\documentclass[10pt,twocolumn,letterpaper]{article}

\usepackage{wacv}
\usepackage{times}
\usepackage{epsfig}
\usepackage{graphicx}
\usepackage{amsmath}
\usepackage{amssymb}
\usepackage{enumitem}
\usepackage{caption}
\usepackage{subcaption}
\usepackage{array}
\usepackage[export]{adjustbox}
\usepackage{enumerate}
\usepackage{enumitem}
\usepackage{url}
\usepackage{xcolor}
\usepackage[export]{adjustbox}
\usepackage{breqn}
\usepackage{adjustbox}
\usepackage{pifont}% http://ctan.org/pkg/pifont
\newcommand{\cmark}{\ding{52}}%
\newcommand{\xmark}{\ding{54}}%
\usepackage{enumitem}
\wacvfinalcopy % *** Uncomment this line for the final submission

\def\wacvPaperID{****} % *** Enter the wacv Paper ID here

\ifwacvfinal\fi
\begin{document}

%%%%%%%%% TITLE
\title{DCIL: Deep Contextual Internal Learning for Image Restoration and Image Retargeting}

\author{Indra Deep Mastan and Shanmuganathan Raman\\
Indian Institute of Technology Gandhinagar\\
Gandhinagar, Gujarat, India\\
{\tt\small \{indra.mastan, shanmuga\}@iitgn.ac.in}}

%% omit the following lines up until the closing ``}''.
%% Additional authors and addresses can be added with ``\and'',
%% just like the second author.
%% To save space, use either the email address or home page, not both
%\and

\maketitle
\ifwacvfinal\thispagestyle{empty}\fi

%%%%%%%%% ABSTRACT
\begin{abstract}
%%%%%%%%%%%%%%%%%%%%%%%%%%%%%%%%%%%% Background %%%%%%%%%%%%%%%%%%%%%%%%%%%%%%%%%%%%    
Recently, there is a vast interest in developing unsupervised methods that are independent of the feature learning from the training data, \textit{e.g.},  deep image prior \cite{Ulyanov2018CVPR}, zero-shot learning \cite{shocher2018zero}, and internal learning \cite{shocher2018internal}. These methods are based on the common goal of maximizing the quality of image features learned from a single image despite inherent technical diversity.
%%%%%%%%%%%%%%%%%%%%%%%%%%%%%%%%%%%% Our work theoretical introduction %%%%%%%%%%%%%%%%%%%%%%%%%%%%%%%%%%%%    
In this work, we bridge the gap between the various unsupervised approaches above and propose a general framework for image restoration and image retargeting. We use contextual feature learning and internal learning to improvise the structure similarity between the source and the target images. 
%%%%%%%%%%%%%%%%%%%%%%%%%%%%%%%%%%%% Targeted Applications %%%%%%%%%%%%%%%%%%%%%%%%%%%%%%%%%%%%    
We perform image resizing application in the following setups: classical image resizing using super-resolution, a challenging image resizing where the low-resolution image contains noise, and content-aware image resizing using image retargeting. We also compare our framework with relevant state-of-the-art methods.
\end{abstract}

%%%%%%%%% BODY TEXT
\section{Introduction}\label{sec: introduction}
Deep learning based supervised models could implicitly capture the image prior by feature learning on a collection of images \cite{ledig2016photo, zhang2017learning,  bigdeli2017image, yang2017high, wang2017high, zhang2018image, anwar2019densely}. However, deep feature learning using training data could suffer from transformation bias or model collapse \cite{zhang2018separating, creswell2018generative}. Recently, there is a vast interest in using a convolutional neural network (CNN) to minimize the use of training samples \cite{Ulyanov2018CVPR, mechrez2018contextual, gandelsman2018double, shocher2018zero, mastan2019multi, shocher2018internal}. More specifically, the following unsupervised models are remarkably successful \cite{shocher2018zero, Ulyanov2018CVPR, shocher2018internal}. Unsupervised models could allow image restoration when the degradation process is complex and/or unknown and obtaining realistic data for supervised training is difficult \cite{Ulyanov2018CVPR}. 

Unsupervised image feature learning attracts various applications such as image super-resolution, inpainting, and image retargeting.  Shocher  \textit{et al.} proposed zero-shot super-resolution (ZSSR) which does not use any training dataset  \cite{shocher2018zero}. Another research thread for training-data independent methods is Deep Image Prior (DIP) proposed by Ulyanov  \textit{et al.} \cite{Ulyanov2018CVPR}. DIP bridges the gap between handcrafted image prior based classical methods and CNN  based deep prior. It shows that the structure of the encoder-decoder network itself works as the image prior. Later, Raman and Mastan gave a generalization of \cite{Ulyanov2018CVPR} and showed various aspects of the relationship between network construction and image restoration \cite{mastan2019multi}. For example, skip connections improve super-resolution but adversely affect image inpainting \cite{mastan2019multi}. 

ZSSR and DIP compute pixel-to-pixel loss (\textit{e.g.}, mean squared error MSE). The pixel-to-pixel loss is limited to the applications which have a  spatial correspondence between the pixels of the source and the target images (aligned image data). Mechrez  \textit{et al.}  have proposed contextual loss for non-aligned image data applications, \textit{e.g.}, style transfer  \cite{mechrez2018contextual, mechrez2018learning}. However, their approach is not completely independent of training samples. We call image feature learning by minimizing the contextual loss as contextual feature learning (CFL). 

Recently, Shocher \textit{et al.} proposed Internal-GAN (InGAN) for image retargeting without using any training samples. Image retargeting requires feature transfer when there is no spatial correspondence between pixels of the source image and the target image (non-aligned image data) \cite{shocher2018internal}. InGAN considers image retargeting as a distribution matching problem to take advantage of GAN. Shocher \textit{et al.} observed that the reconstructions suffer from the object partition ambiguity. 

There is a technical diversity in the unsupervised methods described above. However, they are all subjected to maximizing the quality of image feature learning from a single image. There are two interesting challenges here. (1)  What aspect of the network would help for the task of image generation in the limited contextual understanding due to the lack of feature learning from the training data? (2) What should be the structure of the loss function when the source image and the target image are non-aligned and do not have spatial correspondence?

To better understand the challenges above, let us consider the task of resizing an image. Super-Resolution (SR) scales the entire image, whereas image retargeting resizes the input while preserving the size and the aspect ratio of the local elements \cite{shocher2018internal}. Another image resizing application would be to scale a low-resolution image, which contains noise, termed as Denoising-Super-Resolution (DSR). Image resizing in DSR setting is more general and challenging than image super-resolution as it also removes noise from a low-resolution image. 

The training data independent image resizing in various scenarios described above would require a careful design of network and loss function. DIP and ZSSR perform image restoration using pixel-to-pixel loss. Therefore, they are not applicable to image retargeting (non-aligned image data). InGAN performs image retargeting, but is not studied for DSR setting \cite{shocher2018internal}. 

%It is commonly used in the style transfer application where the one combines style features and the content features of the different image to create a stylized output, e.g., gram loss. 
%This paper investigates the image specific priors learned from a single image. These priors are based on different ways to realize the statistics of the image. An image can be seen as a sample taken from the patch distribution of the natural images. Based on the above insight, the InGAN learns the patch distribution of the natural images. The image transformation is achieved from the contextual loss is performed by realizing the image as a collection of image features (high dimensional points). 
%To develop a method which generalizes to a new target image, one has to minimize the dependence upon the training samples. 

We propose deep contextual internal learning (DCIL) for image retargeting and image restoration. Our models include structure of the network as the implicit image prior and an image degradation based loss term. The desired image is reconstructed by finding the optimal solution of the model. The key idea is to maximize deep feature learning from a single image by a modular network structure with a generalized loss function which works for aligned and non-aligned image data. We use diverse techniques such as deep prior learning, adversarial learning, and CFL. Deep prior learning fits the generator network by maximizing the likelihood of weights given the corrupted image and restoration model. Adversarial learning and CFL perform distribution matching to generate realistic image patches. %The framework is novel as it performs image resizing in various challenging scenarios (\textit{i.e.}, image restoration and image retargeting, Table~\ref{table: compare}). 

There is an interesting contrast to our objectives. On the one hand, we need an image restoration strategy which enhances the image features present in the input corrupted image. On the other end, image retargeting specific reconstruction, which uses similar image features from input image for synthesizing objects in the output image (distribution matching). Image distribution matching between the corrupted image and the output image could adversely influence output image. Therefore, we provide modularity in the network structure with a task-specific loss function (Sec.~\ref{sec: method}). The network provides a high impedance to the noise and allows reconstruction of the signal \cite{Ulyanov2018CVPR}. The loss function influence the quality of learned features by minimizing the dis-similarity in the features of the source image and the output image. 

Network construction in DCIL is based on various network components. These components are as follows: network depth, skip connections, a cascade of network input, and network composition. We do not use cascading of network input as it does not provide a significant enhancement of prior learning \cite{mastan2019multi}. We take advantage of residual blocks as it improves the generator output \cite{Isola_2017_CVPR, zhu2017unpaired}.

We formalize network construction and explain the abstract description of the network to simplify the network design in the presence of diverse components (Sec.~\ref{ssec: netComponents}). %We provide a network description of the generator and the discriminator networks for simplifying the network construction module.

After network construction, DCIL iteratively minimizes the loss between the source image and the target image.  DCIL loss compares image features between the source image and the target image in three ways: pixel-to-pixel, patch-to-patch, and contextual features comparison. The motive behind this loss is to capture better image statistics by comparing the diverse set of image features. (1) Pixel-to-pixel comparison is similar to the MSE based reconstruction loss \cite{Ulyanov2018CVPR, shocher2018internal}. (2) Patch-to-patch comparison is made using the adversarial loss \cite{shocher2018zero, shocher2018internal}. (3) Contextual features comparison between the source and the target image is done using the contextual loss \cite{mechrez2018contextual}. 

Adversarial loss generate realistic samples by preserving the distribution of image patches 
\cite{shocher2018internal}. Contextual loss is motivated to enhance the structural similarity of the objects in the output image \cite{mechrez2018learning}. The reconstruction loss ensures the preservation of global image features in the target image. 

Image resizing using DSR and SR are both naturally occurring. We corrupt the input image to a high degree to observe the quality of image features captured in DIP \cite{Ulyanov2018CVPR} and CFL \cite{mechrez2018contextual}. We show that DCIL generates reconstruction that is comparable to that of the other relevant unsupervised frameworks (Sec.~\ref{SSEC: multiCorr}). Mechrez \textit{et al.} have shown that CFL exhibits natural internal statistics for SR \cite{mechrez2018learning}. We illustrate the performance of CFL in the training data-independent setup for SR task (Sec.~\ref{SSEC: super}). 

DCIL performs image retargeting by changing the size and the shape of the generator output. The target image is subjected to preserve the distribution of image patches. Adversarial loss and contextual loss are well suited for the above task. More specifically, the contextual feature learning with the internal patch distribution learning (InGAN) is observed to preserve good object statistics for image retargeting task (Sec.~\ref{SSEC: retargeting}).

The key contributions of the paper are as follows.
%%%%%%%%%%%%%%%%%%%%%%%%%%%%%%%  WACV %%%%%%%%%%%%%%%%%%%%%%%%%%%%%%%%%%%%
%Through extensive experimental validation on three datasets, including comparisons to several baseline methods using a variety of performance metrics, we verify the effectiveness of our proposed framework.
%In addition, we show the proposed framework can be applied to the fewshot image classification task. 
%By training a classifier on the images generated by our model for the few-shot classes, we are able to outperform a state-of-the-art few-shot classification method that is based on feature hallucination.
 %%%%%%%%%%%%%%%%%%%%%%%%%%%%%%%%%%%%%%%%%%%%%%%%%%%%%%%%%%%%%%%%%%%%%%
\begin{itemize}[leftmargin=*] \setlength\itemsep{0em}
\item[1.] We propose a generalized framework (DCIL) for image resize in various scenarios by coupling an internal learning scheme in a \emph{novel} unsupervised contextual feature learning framework (Sec.~\ref{sec: method} and Table \ref{table: compare}).
\item[2.] We verify effectiveness of DCIL by extensive experimentation for DSR, SR, and image retargeting tasks (Sec.~\ref{sec: application}).
\item[3.]  To the best of our knowledge, we are the first to study CFL in an unsupervised framework for DSR task (Table~\ref{tab: multi}).
\item[4.] DCIL preserves the object structure and alignment in the image retargeting output (Fig.~\ref{fig: objAlign} and Fig.~\ref{fig: objStructure}). We also give ablation studies for understanding various aspects of the loss functions (Sec.~\ref{sec: ablation}).  
\end{itemize}

%%%%%%%%%%%%%%%%%%%%%%%%%%%%%%%%%%%%%%%%%%%%%%%%%%%%%%%%%%%%%%%%%%%%%%
%\begin{figure}
%\includegraphics[width=0.10\textwidth]{images/imagePriorComparision/blue.jpeg}
%\caption{Image prior comparison}
%\label{fig: imagePriorComparision}
%\end{figure}
%%%%%%%%%%%%%%%%%%%%%%%%%%%%%%%%%%%%%%%%%%%%%%%%%%%%%%%%%%%%%%%%%%%%%%
\begin{figure*}[t]
\centering
\includegraphics[width=0.68\textwidth]{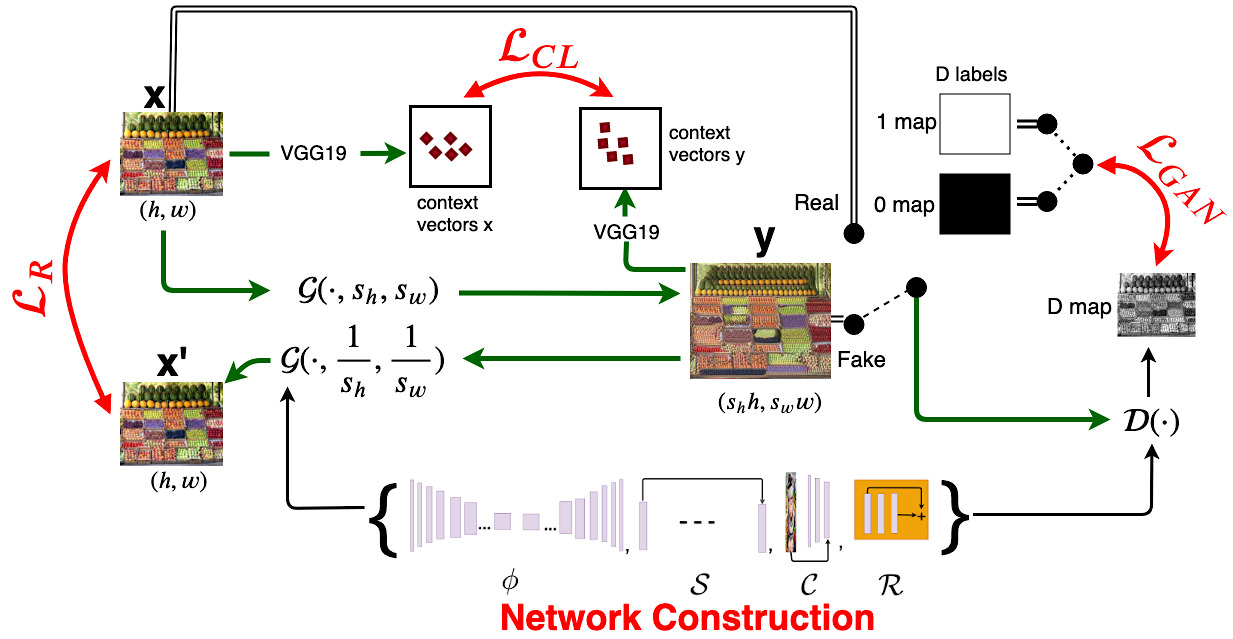}
\caption{\textbf{Deep Contextual Internal Learning (DCIL).} The figure shows the pictorial representation of the DCIL framework for image retargeting. It contains a generator $\mathcal{G}$ that takes an input image $x$ of size $(h, w)$ and outputs an image $y$ of a different size using the scaling factors $(s_h, s_w)$. The output of the generator $y=\mathcal{G}(x, s_h, s_w)$ is fed into the discriminator $\mathcal{D}$ and feature extractor pre-trained $VGG19$ network \cite{simonyan2014very}. The same framework is used for image restoration where the definitions of the loss functions is different. The idea here is to create the generator and discriminator using network construction module and then iteratively minimize the loss functions (we describe various entities of the pictorial representation above in Sec.~\ref{sec: method}). }\label{fig: blockDiagram} \vspace*{-0.5cm}
\end{figure*}%
\begin{figure}\centering
\includegraphics[width=\linewidth]{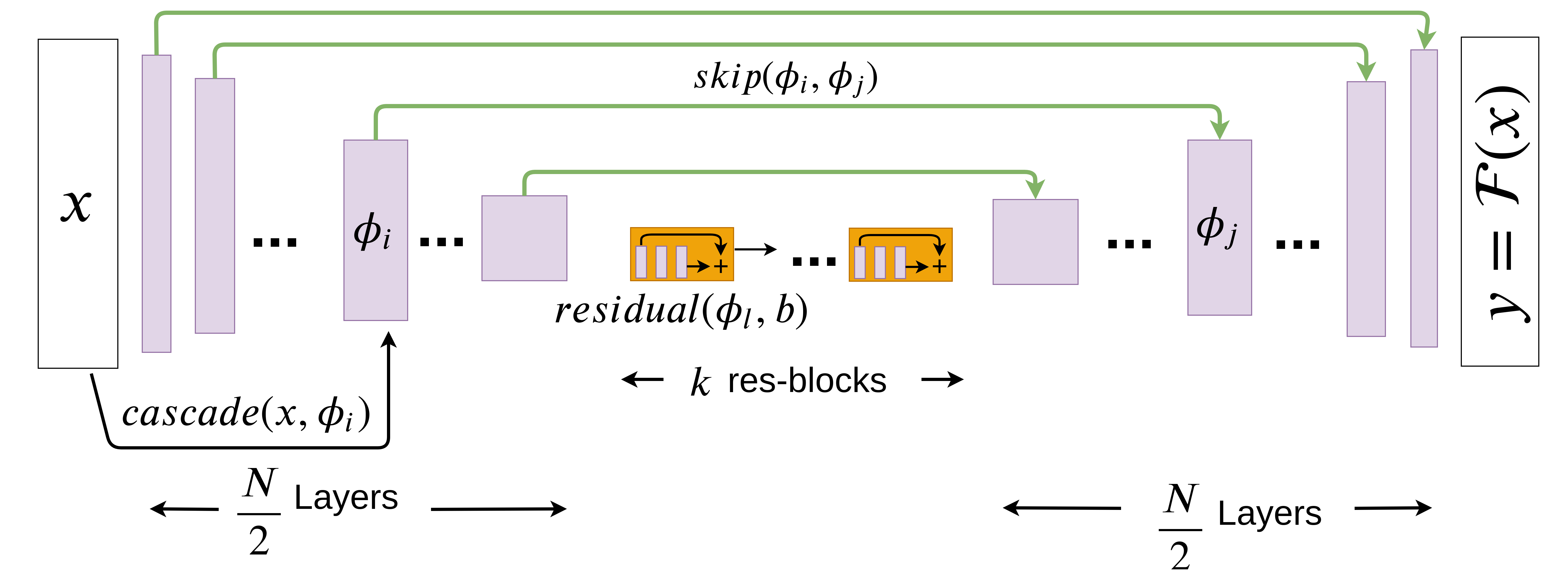}
\caption{\textbf{Network Construction.} The figure illustrate various network components needed for the construction of the network $\mathcal{F}$, defined in Eq.~\ref{eq: networkConstruction} and Sec.~\ref{ssec: netComponents}. $\phi_i$ denotes the network layers. $\text{skip}(\phi_i, \phi_j)$ denote the link between $i^{th}$ and $j^{th}$ layer. $\text{cascade}(x, \phi_i)$ denotes the cascading of input $x$ at $i^{th}$ layer. Similarly, $k$ residual blocks are shown. The network description is $\mathcal{F}=(\Phi, \mathcal{S}, \mathcal{C}, \mathcal{R})$ (Eq.~\ref{eq: networkConstruction}, Eq.~\ref{eq: skip}, and Eq.~\ref{eq: residual}). }\label{fig: networkEx}
\end{figure}

%%%%%%%%%%%%%%%%%%%%%%%%%%%  WACV  %%%%%%%%%%%%%%%%%%%%%%%%%%%%%%%%%%%%%
%InGAN consists ofa Generator G that retargets input x to output y whose size/shape is determined by a geometric transformation T (top left). A multiscale discriminator D learns to discriminate the patch statistics ofthe fake output y from the true patch statistics of the input image (right). Additionally, we take advantage ofG’s automorphism to reconstruct the input back from y using G and the inverse transformation T−1 (bottom left).
%%%%%%%%%%%%%%%%%%%%%%%%%%%%%%%%%%%%%%%%%%%%%%%%%%%%%%%%%%%%%%%%%%%%%%

\section{Related work}\label{sec: related}
%The ability to learn a statistically good image prior promotes various applications \cite{zoran2011learning, levin2007blind, burger2012image, sun2008image, zhang2017image, buades2005non, dabov2007image, efros1999texture, simakov2008summarizing}. The different image priors capture different image statistics \cite{shaham2016visualizing}. For example, Total variation (TV) regularizer \cite{rudin1992nonlinear} captures object boundaries and curvature \cite{bellettini2002total} whereas the self-similarity prior captures the  redundant image features present due to strong internal data repetition in the natural image \cite{irani2009super, zontak2011internal, Ulyanov2018CVPR, shocher2018internal}. 
Our approach is related to the training data-independent CNN based methods. We bridge the gap between various unsupervised methods proposed to minimize the use of training samples such as deep image prior \cite{Ulyanov2018CVPR}, contextual learning \cite{mechrez2018learning}, and internal learning \cite{shocher2018internal}. Unlike the classical image prior \cite{aharon2006k, dong2012nonlocally, gu2017weighted}, the deep prior learning \cite{Ulyanov2018CVPR} shows that hand-crafted structure of the network work as a prior to capture good image statistics for various image restoration tasks.  Zero-shot learning uses the internal recurrence of information inside a single image to collect various image specific statistics for super-resolution \cite{shocher2018zero}.  InGAN uses multi-scale patch discriminator for learning the patch distribution from the source image \cite{shocher2018internal}. DCIL gets network structure insight from \cite{mastan2019multi}. The construction of DCIL loss is related to \cite{Ulyanov2018CVPR, mechrez2018learning, shocher2018internal}.

\section{Deep Contextual Internal Learning}\label{sec: method}
%%%%%%%%%%%%%%%%%%%% WACV %%%%%%%%%%%%%%%%%%%%%
%HIGH LEVEL: Our InGAN is an image-conditional GAN (e.g., [17]) that maps an input image (as opposed to noise) to a remapped target output image. It uses a genera- tor, G, a discriminator, D, and re-uses G for decod- ing/reconstructing the input given the output, as depicted in Fig. 2. 
%
%In detail, our method works as follows. Explaining the components.
%%%%%%%%%%%%%%%%%%%%%%%%%%%%%%%%%%%%%%%%%%%%%
DCIL uses image-conditional GAN to map the input image (as opposed to noise) to a different size target image. It performs the image restoration and retargeting without using training data-set.  Fig.~\ref{fig: blockDiagram} shows how the major parts of the DCIL framework (highlighted in red) are connected. We describe how these components are related as follows. \\

\noindent \textbf{Overview.} Given a source image $x$, the objective is to output a target image $y$ from the target distribution $Y$. The learning procedure is unsupervised and only uses information from the source $x$. The image in the target domain is of different size than that of $y$ (image resizing). For example, in Denoising-Super-Resolution (DSR), the source image $x$ is a low resolution noisy image and the target domain $Y$ is a set of high-resolution clean image (image restoration). 

DCIL constructs generator $\mathcal{G}$ and discriminator $\mathcal{D}$ using the network components described in Sec.~\ref{ssec: netComponents}. The network parameters are randomly initialized. The source image $x$ is fed to the generator $\mathcal{G}(\cdot, s_h, s_w)$, where $s_h$ and $s_w$ are the scaling factors for height and width. Next, we iteratively minimize the total loss (Eq.~\ref{eq: DCIL}) computed between the generator output $y=\mathcal{G}(x, s_h, s_w)$ (\textit{i.e.}, target image) and input source image $x$. The total loss consists of contextual loss $\mathcal{L}_{CL}$ for contextual feature learning, adversarial loss $\mathcal{L}_{GAN}$ for internal patch distribution learning, and reconstruction loss $\mathcal{L}_{R}$ for global features learning. We describe these major parts of the DCIL below.

\subsection{Network Construction} \label{ssec: netComponents}
We simplify the network construction based on the major components and abstract out network layer specific details. The structure of the network influences the quality of the image features captured. More specifically, the network structure itself works as a prior in the training data-independent methods \cite{Ulyanov2018CVPR}. Therefore, the purpose of this component is to provide more modularity and a degree of freedom for the DCIL framework. 

Consider a network $\mathcal{F}$ which could be a generator or a discriminator. The formal description of network $\mathcal{F}$ is given in Eq.~\ref{eq: networkConstruction}.  
\begin{equation}\label{eq: networkConstruction}
\mathcal{F} = \big(\Phi,\mathcal{S},\mathcal{C},\mathcal{R}\big)
\end{equation}
Here, $\phi$ denotes the set of network layers, $\mathcal{S}$ denotes the configurations of the skip connections, $\mathcal{C}$ denotes the set of layers for which the cascading of the network inputs is performed, and $\mathcal{R}$ denotes the residual blocks.  We discuss the major network components given in Eq.~\ref{eq: networkConstruction} as follows.
\begin{itemize}[leftmargin=*]\setlength\itemsep{0em}
\item \underline{\textit{Network Layers ($\Phi$)}}. The network layers store image representations. Given a network  $\mathcal{F}$ with depth $N$, let $\Phi = \{\phi_l \}_{l=1}^N$ be the set of layers present in the network. Here, $\phi_i$ could be a convolution layer, an activation layer, or a batch normalization layer.
\item \underline{\textit{Skip connections ($\mathcal{S}$)}}. The skip link between the layers $\phi_i$ and $\phi_j$, where $i<j$, is made by concatenating the output of the layer $\phi_{j-1}$ with the output of the layer $\phi_i$ and then feeding into the layer $\phi_j$. Let $\text{skip}(\phi_i, \phi_j) = \text{conv}(\text{conv}(\phi_i) \| \phi_{j-1})$ denote the skip link between the layers $\phi_i$ and $\phi_j$. The set of skip connections of $\mathcal{F}$ is given in Eq.~\ref{eq: skip}.
\begin{equation}\label{eq: skip}
\mathcal{S} = \{\text{skip}(\phi_i, \phi_j): \phi_i, \phi_j \in \phi; i<j  \}
\end{equation}
To simplify the network description, let us denote $\mathcal{S}$ by the set of tuples where each tuple contains the network layer identifier for the skip connection. For example, $\mathcal{S}= \{(1, N), (2, N-1) \}$ denotes the two skip connections, $\text{skip}(\phi_1, \phi_N)$ and $\text{skip}(\phi_2, \phi_{N-1})$.
\item \underline{\textit{Cascading of network input ($\mathcal{C}$)}}. It is a procedure to successively resize the network input $x$ and then feed it into the intermediate layer $\phi_i$ of the network.  We have not used cascading of network input in DCIL as it was shown to not significantly improve the performace \cite{mastan2019multi}. We denote this by $\{ \}$ in the network description for completeness. We have described it more in the supplementary material. 
%Let $cascade(\phi_i) = conv(\mathcal{R}(x, \phi_i.size)\|\phi_{i-1})$ denotes the cascading of the network input $x$ into the layer $\phi_i$, where $\mathcal{R}$ denotes the resize operation. The cascading of network input at different layers is denoted in Eq.~\ref{eq: cascade}. 
%\begin{equation}\label{eq: cascade}
%\mathcal{C} = \{cascade(x, \phi_i): \phi_i \in \Phi\}
%\end{equation}
%To simplify the network description, let us denote $\mathcal{C}$ by the set of target layers, \textit{e.g.}, $\mathcal{C}= \{2, 4\}$ denotes cascading of network inputs at $\phi_2$ and $\phi_4$.
\item \underline{\textit{Residual Block ($\mathcal{R}$)}}. The residual learning framework helps in training higher depth networks while preventing the vanishing gradients problem \cite{he2016deep}. It adds the output of two convolution layers $b$ blocks apart. Let $\text{residual}(\phi_l, b) = \text{add}(\text{conv}(\phi_{l+b}), \phi_l)$ denotes the output residual block $\{\phi_i \}_{i=l+1}^{l+b}$ of length $b$. The set of residual blocks $\mathcal{R}$ of $\mathcal{F}$ is defined in Eq.~\ref{eq: residual}. 
\begin{equation}\label{eq: residual}
\mathcal{R} = \{\text{residual}(\phi_l, b): \phi_l \in \Phi, b\in [N] \}
\end{equation}
To simplify the description, let us denote $\mathcal{R}=[k]$, where $k$ is the number of residual blocks.
\end{itemize}

\noindent \textbf{Generator.}\label{ssec: gen} Fig.~\ref{fig: networkEx} shows an example construction of the generator $\mathcal{G}: X \rightarrow Y$  to show the four-tuple network description. It is an encoder-decoder network which maps the given source image $x$ to the target image $y=\mathcal{G}(x, s_h, s_w)$, $x \sim X$ and $y \sim Y$. The network description of the generator $\mathcal{G}=\big(\Phi, \mathcal{S}, \mathcal{C}, \mathcal{R}\big)$ based on Eq.~\ref{eq: networkConstruction} is defined in Eq.~\ref{eq: generator}. 
\begin{equation}\label{eq: generator}
\mathcal{G} = \Big( \{\phi_l \}_{l=1}^{N}, \{(i, N-i)\}_{i=2}^{\frac{N}{2}-1}, \{ \}, [k] \Big)
\end{equation}
Here, $\Phi = \{\phi_l \}_{l=1}^{N}$ is the set of network layers. We have defined network components $\mathcal{S}$ and  $\mathcal{R}$ in Eq.~\ref{eq: skip} and Eq.~\ref{eq: residual}. $\{ \}$ denotes that cascading of the network input is not performed. We use network configurations defined in Eq.~\ref{eq: superN} and Eq.~\ref{eq: retargetingN} for our experiments. \\

\noindent \textbf{Discriminator.}\label{ssec: dis} It maps the generated image $\mathcal{G}(x) \in Y$ to a patch discriminator $m \in M$, where each entry in $m$ denotes the probability of a patch coming from the patch distribution of the natural image, \textit{i.e.}, $\mathcal{D}: Y \rightarrow M$. We define discriminator as $\mathcal{D}(z) =  \sum_{i=1}^{4} w_i D^i(z)$. 
%\begin{equation}\label{eq: discriminatorSum}
%\mathcal{D}(z) =  \sum_{i=1}^{4} w_i D^i(z)
%\end{equation}
Here, each $D^i$ is a convolution patch discriminator which outputs a map containing the scores of the image patches to be real. And there are four discriminators. The description of discriminator $D^i$ is given in Eq.~\ref{eq: discriminator}. 
\begin{equation}\label{eq: discriminator}
D^i = \Big( \{d^i_l \}_{l=1}^{4}, \{ \}, \{ \}, \{ \} \Big)
\end{equation}
Here, $D^i$ is a CNN with 4-layers. The empty set $\{ \}$ denote the absence of the network component. Therefore, $D^i$ does not have skip links, no residual blocks, and no cascading of network input.  The multiscale discriminator $\mathcal{D}$ matches the patch distribution over a range of patch sizes capturing both the fine-grained details as well as the coarse structures in the image \cite{shocher2018internal} \footnote{In the supplementary material, we provide more details on the generator and the discriminator network.}. 

%A factor of $\sqrt{2}$ downsamples the scale of the discriminator from one scale to another (\textit{i.e.}, $D^i$ to $D^{i+1}$). Weights are not shared between the different scale discriminators.
%The discriminator is a multi-scale discriminator. Training D with LSGAN often leads to mode collapse where retargeted y. The multiscale patches of y are drawn from the input image's distribution, i.e., important visual information is missing from the retargeted y. 
%To create perceptually good images, the discriminator should be able to discriminate the output of the generator and real images. One way is to have a large receptive field using high depth network and large convolutional kernels. However, this introduces high model capacity and vanishing gradients problem. 
%We compute the least-squares of the discriminator as it is observed to be more stable during the training \cite{mao2017least}. 
\subsection{Loss Function} \label{ssec: lossFunctions}
Given the source image $x$, the objective is to generate the target image $\mathcal{G}(x)=y$ from the target domain $Y$. Total loss function $\mathcal{L}$ minimizes the difference in features of the source image and the target image at different feature representations: pixel-to-pixel comparison (reconstruction loss $\mathcal{L}_{R}$), context vectors comparison (contextual loss $\mathcal{L}_{CL}$), and patch-based comparison (adversarial loss $\mathcal{L}_{GAN}$). The total loss function $\mathcal{L}$ is described in Eq.~\ref{eq: DCIL}.
\begin{equation}\label{eq: DCIL} 
\begin{split}
\mathcal{L} = \lambda_{\mathcal{C}} \; \mathcal{L}_{CL}(\mathcal{G}(x), x) + \lambda_{\mathcal{G}} \; \mathcal{L}_{GAN}(\mathcal{G},\mathcal{D} ,x,y) \\ + \lambda_{\mathcal{R}} \; \mathcal{L}_{R}(\mathcal{G}, \mathcal{D}, x, y) 
\end{split}
\end{equation}
Here, $\mathcal{G}$ and $\mathcal{D}$ are both CNN described in Sec.~\ref{ssec: netComponents}.  The terms $\lambda_{\mathcal{C}}$, $\lambda_{\mathcal{G}}$, and $\lambda_{\mathcal{R}}$ are the coefficients of the loss functions.

The intuition behind loss in Eq.~\ref{eq: DCIL} is that minimizing feature differences at different image representations could help in maximizing the image feature learning from the source image. 
$\mathcal{L}_{CL}$ compares context vectors to make the distribution of the generator output to be contextually similar to the distribution of the natural images \cite{mechrez2018contextual}. $\mathcal{L}_{GAN}$ is aimed to output distribution of the image patches, which is indistinguishable from the patch distribution of the natural images. $\mathcal{L}_{R}$ performs the pixel-to-pixel comparisons between the source image and the target image or an inverse mapping of the target image. It ensures that we do not miss any of the object details in the generator output image. We now describe the loss terms used in Eq.~\ref{eq: DCIL} for completeness. \\

\noindent \textbf{Contextual loss ($\mathcal{L}_{CL}$).} It is used to enhance the contextual features in the reconstruction. The set of context vectors are obtained by feeding image $x$ and $y$ into pre-trained VGG19 $\phi$ \cite{simonyan2014very}. In Fig.~\ref{fig: blockDiagram}, we have pictorially shown context vectors as the output of VGG19. Let $\phi^l(x)$ and $\phi^l(y)$ denote the feature extracted from layer $l$ of the network $\phi$. The contextual loss is defined in Eq.~\ref{eq: context}. 
\begin{equation}\label{eq: context}
\mathcal{L}_{CL}(x, y, l) = - \log CX(\phi^l(x),\phi^l(y)) 
\end{equation} 
Here, $CX$ denotes the contextual similarity measure. It is computed by considering cosine distance between the context vectors extracted from the network $\phi$ \cite{mechrez2018contextual}. Eq.~\ref{eq: context} minimizes dissimilarities between the contextual feature computed from the source image $x$ and the target image $y$.  CX is  normalized and lies in the range $[0,1]$. \\
%The segmentation mask is used to fill in the missing regions in the image by computing the contextual loss between the missing region of the network output and the corrupted image. 
%The key aspect is that the images in the comparison do not need to have the feature alignment and the same size. 

\noindent \textbf{Adversarial loss ($\mathcal{L}_{GAN}$).} The purpose of adversarial learning is to synthesize new image features in the output image from the patch distribution of the natural images. It is a sum of the generator loss $\mathcal{L}_{\mathcal{G}}$ and the discriminator loss $\mathcal{L}_{\mathcal{D}}$. The generator ${\mathcal{G}}$ and the discriminator ${\mathcal{D}}$ are both CNN. The generator loss $\mathcal{L}_{\mathcal{G}}$ and the discriminator loss $\mathcal{L}_{\mathcal{D}}$ are used for distribution matching. $\mathcal{G}$ generates the desired image. $\mathcal{D}$ tries to distinguish the output of $\mathcal{G}$ and the source image. Therefore, the generator learns the patch distribution through the interaction with the discriminator. We show the adversarial loss in Eq.~\ref{eq: ganLoss}. 
\begin{equation}\label{eq: ganLoss}
\mathcal{L}_{GAN}({\mathcal{G}}, {\mathcal{D}}, x, y) = \mathcal{L}_{\mathcal{G}}(x) + \mathcal{L}_{\mathcal{D}}(x, y)
\end{equation}
Here, $\mathcal{G}$ outputs the target image $\mathcal{G}(x)=y$. 
The feature learning in the adversarial framework could suffer from mode collapse \cite{arjovskyB17, arora2017generalization}. The use of multi-scale discriminator prevents it by maximizing feature learning by comparing the reconstruction at multiple scales \cite{shocher2018internal}. We have described the multi-scale discriminator in Sec.~\ref{ssec: netComponents}.\\ 
 
\noindent \textbf{Reconstruction loss ($\mathcal{L}_{R}$).} It is used to maximize the likelihood of randomly initialized network weights. One could define a spatial correspondence in the case of image restoration \cite{Ulyanov2018CVPR}. However for image retargeting, reconstruction loss in a cycle consistent approach performs well as the generator output does not have spatial correspondence with the source image \cite{shocher2018internal}. Therefore, the two different ways of computation of reconstruction loss are as follows.

$\mathcal{L}_{R}$ for image restoration is computed between the generator output $\mathcal{G}(x)$  and the source image $x$, as in Eq.~\ref{eq: reconstruct1}.
\begin{equation}\label{eq: reconstruct1}
\mathcal{L}_{R}({\mathcal{G}}, s_h, s_w, x, y) = \| {\mathcal{G}}(x, s_h, s_w) - x) \|
\end{equation}
$\mathcal{L}_{R}$ for image retargeting is computed between the source image $x$ and the inverse mapping of the generator output ${\mathcal{G}}(y)$, where $y={\mathcal{G}}(x, s_h, s_w)$, as in Eq.~\ref{eq: reconstruct2}. 
\begin{equation}\label{eq: reconstruct2}
\mathcal{L}_{R}({\mathcal{G}}, s_h, s_w, x, y) = \| {\mathcal{G}}(y, \frac{1}{s_h}, \frac{1}{s_w}) - x) \|
\end{equation}

%In Fig.~\ref{fig: blockDiagram}, the output $G(x)$ is used to compute reconstruction loss $\mathcal{L}_{R}$. 
%In other words, the automorphism of $G$  and re-use $G$ (i.e., $G(y)$) to reconstruct $x$ back from the retargeted image $y$ is performed. This is to prevent the mode collapse.  

\section{Applications}\label{sec: application}
%%%%%%%%%%%%%%%%%%%%%%%%%%%%% WACV %%%%%%%%%%%%%%%%%%%%%%%%%%%%
%We now show experimentally how the proposed prior works for diverse image reconstruction problems. Due to space limitations, we present a few examples and numbers and include many more in the Supplementary material and the project webpage [30].
%%%%%%%%%%%%%%%%%%%%%%%%%%%%%%%%%%%%%%%%%%%%%%%%%%%%%%%%%%%%%

In this section, we describe image resizing in two different setups. The first setup is image restoration problems. There are two ways for it. DSR where the low-resolution input contains noise. SR where the low-resolution input does not contain noise. The second setup for image resizing is content-aware image retargeting. We describe these applications below.

%%%%%%%%%%%%%%%%%%%%%%%%%%%%%%%%%%%%%%%%%%%%%%%%%%%%%%%%%%%%%%%%%%%%%%%
%\subsection*{Super-resolution.}
%\begin{itemize}
%\item Three groups of the loss functions. MSE, Contextual, and GAN. 
%\item 
%\end{itemize}
%\begin{subfigure}[b]{0.22\textwidth}
%\includegraphics[width=\linewidth]{images/super/img_001_SRF_4_HR.jpg}
%\end{subfigure}
%\begin{subfigure}[b]{0.22\textwidth}
%\includegraphics[width=\linewidth]{images/super/img_001_SRF_4_LR.jpg}
%\end{subfigure}
%\begin{subfigure}[b]{0.22\textwidth}
%\includegraphics[width=\linewidth]{images/super/S_img_001_SRF_4_HR.jpg}
%\end{subfigure}
%\begin{subfigure}[b]{0.22\textwidth}
%\includegraphics[width=\linewidth]{images/super/superResolution__img_001_SRF_4_HRpng__lr_1e-05_1e-05_00001_1e-05_v_l_00001_g_l_001_m_l_10_c_l_0001png.jpg}
%\end{subfigure} \\
%\begin{subfigure}[b]{0.22\textwidth}
%\includegraphics[width=\linewidth]{images/super/img_002_SRF_4_HR.jpg}
%\end{subfigure}
%\begin{subfigure}[b]{0.22\textwidth}
%\includegraphics[width=\linewidth]{images/super/img_002_SRF_4_LR.jpg}
%\end{subfigure}
%\begin{subfigure}[b]{0.22\textwidth}
%\includegraphics[width=\linewidth]{images/super/S_img_002_SRF_4_HR.jpg}
%\end{subfigure}
%\begin{subfigure}[b]{0.22\textwidth}
%\includegraphics[width=\linewidth]{images/super/superResolution__img_002_SRF_4_HRpng__lr_1e-05_1e-05_00001_1e-05_v_l_00001_g_l_001_m_l_10_c_l_0001png.jpg}
%\end{subfigure} \\
%%%%%%%%%%%%%%%%%%%%%%%%%%%%%%%%%%%%%%%%%%%%%%%%%%%%%%%%%%%%%%%%%%%%%%%
\begin{figure*}\centering
\begin{subfigure}[b]{0.16\linewidth}\begin{center}
\includegraphics[width=\linewidth]{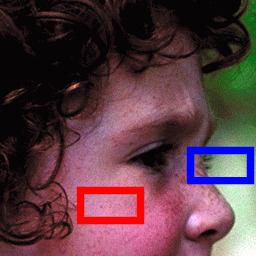} \\
\includegraphics[width=0.46\linewidth]{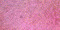}
\includegraphics[width=0.46\linewidth]{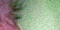}
\end{center} \end{subfigure}
\begin{subfigure}[b]{0.16\linewidth}\begin{center}
\includegraphics[width=\linewidth]{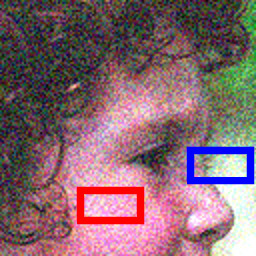} \\
\includegraphics[width=0.46\linewidth]{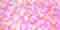}
\includegraphics[width=0.46\linewidth]{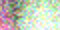}
\end{center} \end{subfigure}
\begin{subfigure}[b]{0.16\linewidth}\begin{center}
\includegraphics[width=\linewidth]{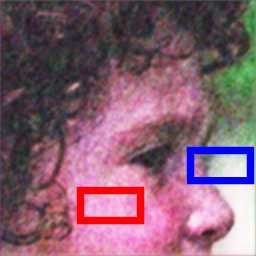} \\
\includegraphics[width=0.46\linewidth]{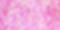}
\includegraphics[width=0.46\linewidth]{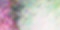}
\end{center} \end{subfigure}
\begin{subfigure}[b]{0.16\linewidth}\begin{center}
\includegraphics[width=\linewidth]{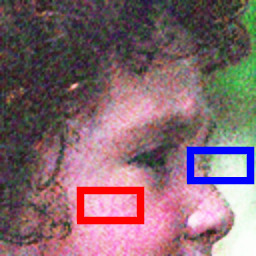} \\
\includegraphics[width=0.46\linewidth]{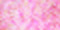}
\includegraphics[width=0.46\linewidth]{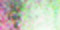}
\end{center} \end{subfigure}
\begin{subfigure}[b]{0.16\linewidth}\begin{center}
\includegraphics[width=\linewidth]{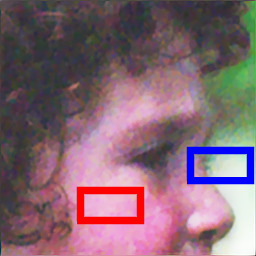} \\
\includegraphics[width=0.46\linewidth]{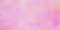}
\includegraphics[width=0.46\linewidth]{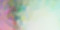}
\end{center} \end{subfigure}\\
\begin{subfigure}[b]{0.16\linewidth}\captionsetup{justification=centering}\begin{center}
\includegraphics[width=\linewidth]{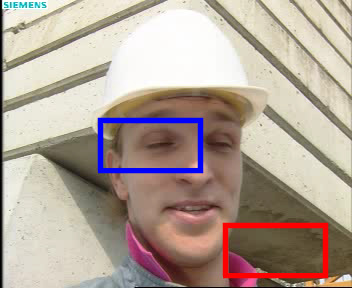} \\
\includegraphics[width=0.46\linewidth]{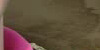}
\includegraphics[width=0.46\linewidth]{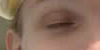}
\end{center}  \vspace*{-0.2cm} \caption{HR image} \end{subfigure}
\begin{subfigure}[b]{0.16\linewidth}\captionsetup{justification=centering}\begin{center}
\includegraphics[width=\linewidth]{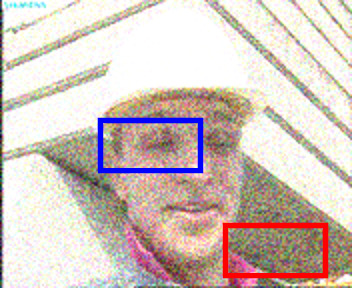} \\
\includegraphics[width=0.46\linewidth]{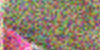}
\includegraphics[width=0.46\linewidth]{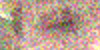}
\end{center}  \vspace*{-0.2cm} \caption{Corrupted image} \end{subfigure}
\begin{subfigure}[b]{0.16\linewidth}\captionsetup{justification=centering}\begin{center}
\includegraphics[width=\linewidth]{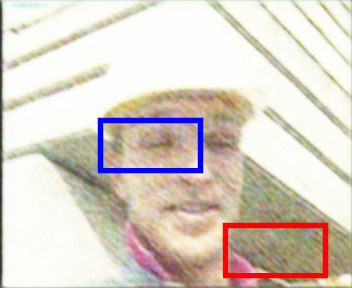} \\
\includegraphics[width=0.46\linewidth]{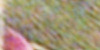}
\includegraphics[width=0.46\linewidth]{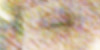}
\end{center}  \vspace*{-0.2cm} \caption{CL \cite{mechrez2018learning}} \end{subfigure}
\begin{subfigure}[b]{0.16\linewidth}\captionsetup{justification=centering}\begin{center}
\includegraphics[width=\linewidth]{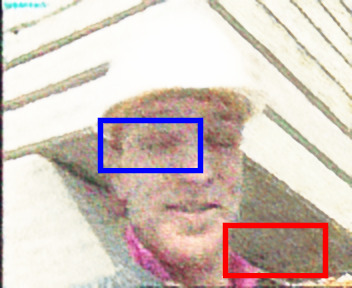} \\
\includegraphics[width=0.46\linewidth]{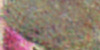}
\includegraphics[width=0.46\linewidth]{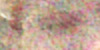}
\end{center}  \vspace*{-0.2cm} \caption{DIP \cite{Ulyanov2018CVPR}} \end{subfigure}
\begin{subfigure}[b]{0.16\linewidth}\captionsetup{justification=centering}\begin{center}
\includegraphics[width=\linewidth]{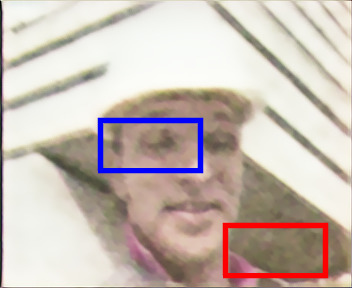} \\
\includegraphics[width=0.46\linewidth]{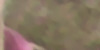}
\includegraphics[width=0.46\linewidth]{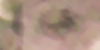}
\end{center}  \vspace*{-0.2cm} \caption{DCIL (Ours)} \end{subfigure} \vspace*{-0.2cm}
\caption{\textbf{2$\times$Denoising-Super-Resolution.} The corrupted low resolution images contain noise with strength $\sigma=100$. CL \cite{mechrez2018learning} and DIP \cite{Ulyanov2018CVPR} create noisy spots in the image restoration output. DCIL (ours) output clean images compared to CL \cite{mechrez2018learning}, DIP \cite{Ulyanov2018CVPR} (see the cropped images below the figures).}\label{fig: MultiCorruption}
\end{figure*}%

%%%%%%%%%%%%%%%%%%%%%%%%%%%%%%%%%%%%%%%%%%%%%%%%%%%%%%%%%%%%%%%%%%%%%%%
%\begin{figure} \centering
%\begin{subfigure}[b]{0.22\linewidth}
%\includegraphics[width=\linewidth]{images/super/img_003_SRF_4_HR.jpg}\caption{HR image}
%\end{subfigure}
%\begin{subfigure}[b]{0.22\linewidth}
%\includegraphics[width=\linewidth]{images/super/img_003_SRF_4_LR.jpg}\caption{LR image}
%\end{subfigure}
%\begin{subfigure}[b]{0.22\linewidth}
%\includegraphics[width=\linewidth]{images/super/S_img_003_SRF_4_HR.jpg}\caption{DIP \cite{Ulyanov2018CVPR}}
%\end{subfigure}
%\begin{subfigure}[b]{0.22\linewidth}
%\includegraphics[width=\linewidth]{images/super/superResolution__img_003_SRF_4_HRpng__lr_1e-05_1e-05_00001_1e-05_v_l_00001_g_l_001_m_l_10_c_l_0001png.jpg}\caption{DCIL}
%\end{subfigure}  \vspace*{-0.2cm}
%\caption{\textbf{4$\times$Super-Resolution.} The perceptual quality comparison for SR. One could observe that DCIL output image is comparable to DIP \cite{Ulyanov2018CVPR} (the images are best viewed after zooming).} \label{fig: super}
%\end{figure}
\begin{figure}[h]
\begin{subfigure}[b]{0.22\linewidth}
\includegraphics[width=\linewidth]{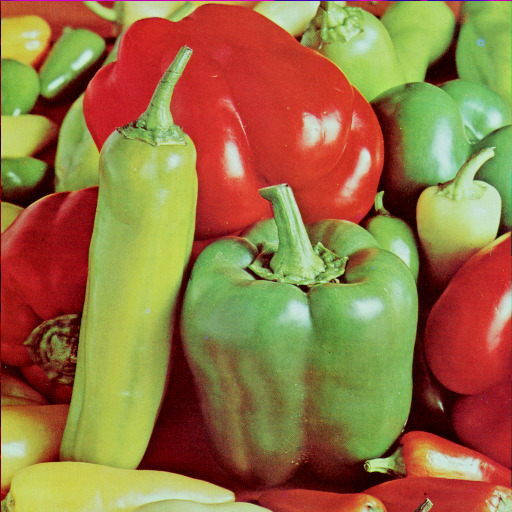} \caption{HR\\ image}
\end{subfigure}
\begin{subfigure}[b]{0.22\linewidth}
\includegraphics[width=\linewidth]{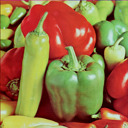} \caption{LR\\ image}
\end{subfigure}
\begin{subfigure}[b]{0.22\linewidth}
\includegraphics[width=\linewidth]{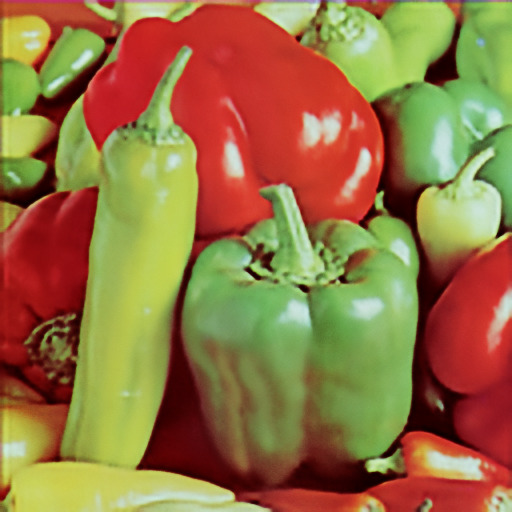} \caption{DIP \cite{Ulyanov2018CVPR}\\ (0.88, 28.2)}
\end{subfigure}
\begin{subfigure}[b]{0.22\linewidth}
\includegraphics[width=\linewidth]{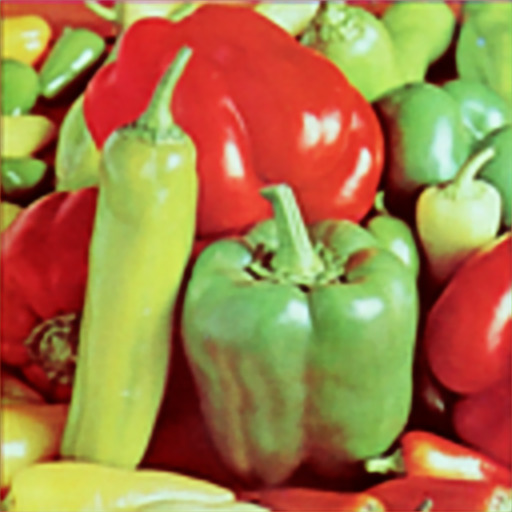} \caption{DCIL\\ (0.88, 24.55)}
\end{subfigure} \vspace*{-0.2cm}
\caption{$4\times$ SR comparison for (SSIM, PSNR) values. DCIL output image is comparable to DIP \cite{Ulyanov2018CVPR} (the images are best viewed after zooming). It could be observed that a higher PSNR value does not imply a higher perceptual quality \cite{mastan2019multi}. }\label{fig: super}
\end{figure}

\begin{table}\begin{center} { \small 
\begin{tabular}{|c|c|c|c|} \hline
& CL \cite{mechrez2018learning} & DIP \cite{Ulyanov2018CVPR} & DCIL (ours) \\ \hline
\textbf{BSD100} & 0.60 & 0.62 & \textbf{0.63} \\ \hline
\textbf{SET14} & 0.62 & 0.66 & \textbf{0.67} \\ \hline
\textbf{SET5} & 0.64 & 0.66 & \textbf{0.66} \\ \hline
\end{tabular}} \end{center} \vspace*{-0.5cm}
\caption{\textbf{2$\times$Denoising-Super-Resolution.} Performance comparision  (SSIM) for 2$\times$SR where low resolution image contains noise with strength $\sigma=100$. }\label{tab: multi}
\end{table}
\subsection{Denoising-Super-Resolution.} \label{SSEC: multiCorr} 
DSR makes the image resize operation challenging as one has to perform two tasks - image denoising and image super-resolution. The description of the generator network for DSR is given in Eq.~\ref{eq: superN}.

\begin{figure*}\centering
\includegraphics[width=\linewidth]{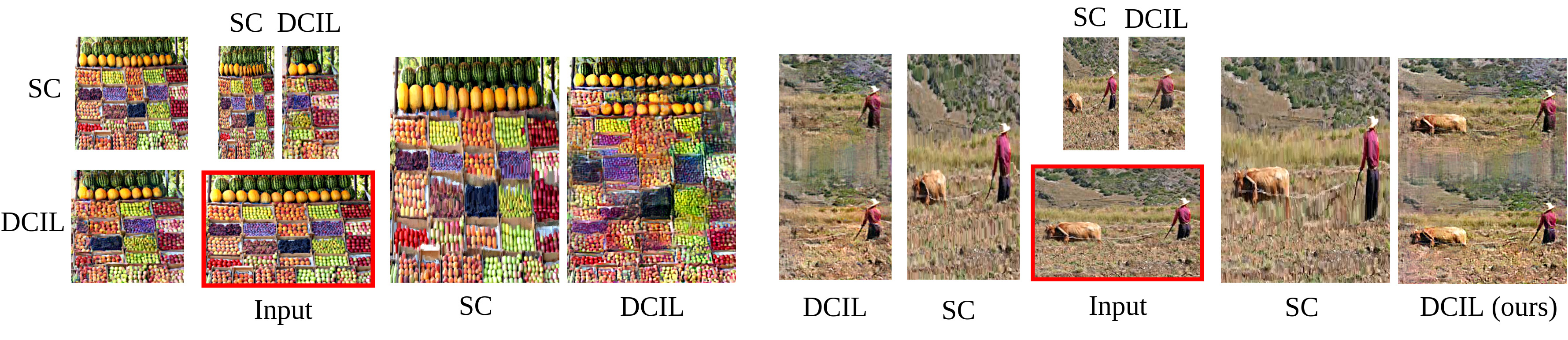} \vspace*{-0.7cm} \caption{\textbf{Image Retargeting (Object Structure).} The size of the local objects (\textit{e.g.}, fruits and man) confirms the preservation of object structure in the image retargeting output. SC \cite{avidan2007seam} does not preserve the structure of the objects (e.g., the man in the 8th column is deformed). DCIL (ours) preserve the structure of the objects by adding new objects or removing objects.}\label{fig: objStructure} 
\end{figure*}%
\begin{figure}\centering
\includegraphics[width=0.87\linewidth]{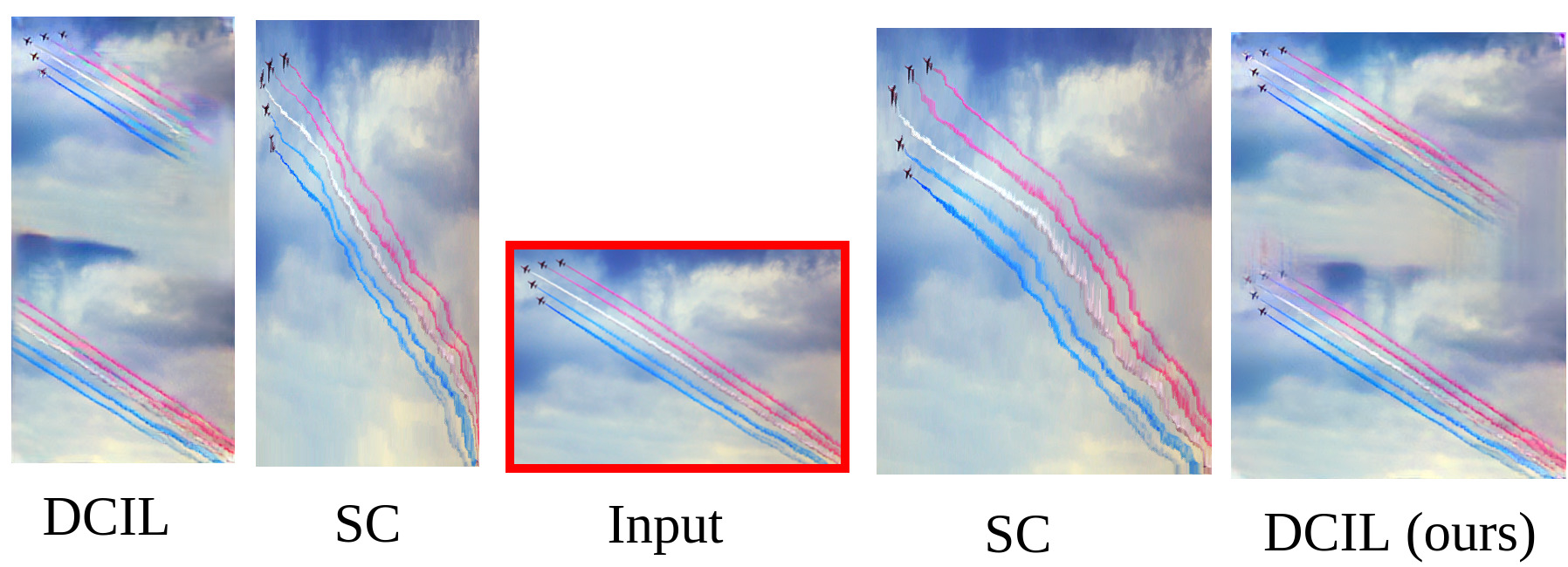}
\vspace*{-0.2cm} \caption{\textbf{Image Retargeting (Object Alignment).} The line-shaped clouds (contrails) produced by aircraft confirms the preservation of the object alignment in the image retargeting output. SC \cite{avidan2007seam} does not preserve the alignment of the objects. DCIL (ours) preserves the alignment of the contrails when increasing height and when increasing width.}\label{fig: objAlign}
\end{figure}

\begin{equation}\label{eq: superN}
{\mathcal{G}_1} = \Big( \{G_l \}_{l=1}^{10}, \{(i, 10-i)\}_{i=2}^{4}, \{\}, \{\} \Big)
\end{equation}
Here, $\{G_l \}_{l=1}^{10}$ is the depth-5 encoder-decoder network where $\{G_l \}_{l=1}^{5}$ are the layers of encoder and $\{G_l \}_{l=6}^{10}$ are the layers of decoder. There are skip connections from encoder layers to the decoder layer in ${\mathcal{G}_1}$. The cascading of the network input is not performed and there are no residual blocks.  Encoder-decoder architecture  contermeasures the mode collapse and improves stability \cite{shocher2018internal}.

Given low-resolution noisy image $\hat{I}$, the loss function for DSR is defined in Eq.~\ref{eq: multiL}.
\begin{equation}\label{eq: multiL}
\begin{split}
\mathcal{L}_1 = \lambda_{\mathcal{C}} &\; \mathcal{L}_{CL}({\mathcal{G}_1}(x), y) + \lambda_{\mathcal{G}_1} \; \mathcal{L}_{GAN}({\mathcal{G}_1},{\mathcal{D}},\hat{I},x) \\ + & \lambda_{\mathcal{R}} \; \| {\mathcal{G}_1}(x) - U_t(\hat{I})) \| +  \lambda_{TV} \|TV(\mathcal{G}_1(x))\|
\end{split}
\end{equation}
Here, $\mathcal{D}$ is similar to the one defined in Eq.~\ref{eq: discriminator}. $U_t(\cdot)$ is the up-sampling operator with the scaling factor as $t$. $\lambda_{TV}$ is the coefficient of the Total variation (TV) regularization. TV norm in Eq.~\ref{eq: multiL} reduces the noise from the corrupted image\footnote{Total variation is a sum of the absolute differences of neighboring pixel values in the input image. It measures the noise in the image.}. We have discussed the loss terms of Eq.\ref{eq: multiL} in Sec.~\ref{ssec: lossFunctions}.

The adversarial loss $\mathcal{L}_{GAN}$ uses the multi-scale patch discriminator to learn the image features at different resolutions. Intuitively, it utilizes the patch replication across multiple scales to augment feature learning. The contextual loss $\mathcal{L}_{CL}$ improvises feature learning at the scale of the target image using context vectors. The reconstruction loss  $\| {\mathcal{G}_1}(x) - U_t(\hat{I})) \|$ provides the global features in the resulting output. 

In Table~\ref{tab: multi}, we give quantitative comparisons for DSR. The aim is to perform  2$\times$SR with denoising, where noise strength is $\sigma=100$. The visual comparison for the generated images is provided in Fig.~\ref{fig: MultiCorruption}. One could observe that we outperform the state-of-the-art methods which we compare with\footnote{We use original implementation of contextual loss (\url{github.com/roimehrez/contextualLoss}) and DIP (\url{github.com/DmitryUlyanov/deep-image-prior}). We generated DIP output images using the default hyper-parameters \cite{Ulyanov2018CVPR}.}. 

%%%%%%%%%%%%%%%%%%%%%%%%%%%%%%%%%%%%%%%%%%%%%%%%%%%%%%%%%%%%%%%%%%%%%%%%
%\begin{table}\begin{center} { \small 
%\begin{tabular}{|c|c|c|c|c|} \hline
%& CL \cite{mechrez2018learning} 0.64 & ZSSR \cite{shocher2018zero} 0.72 &  DIP \cite{Ulyanov2018CVPR} 0.79 & DCIL 0.76\\ \hline
%\textbf{SSIM} & 0.64 & 0.72 & 0.79 & 0.76 \\ \hline
%\end{tabular}} \end{center}  \vspace*{-0.5cm}
%\caption{\textbf{4$\times$ image super-resolution.} Performance comparison on BSD100 data set.
%}\label{tab: super}
%\end{table}
%\begin{figure*}\centering
%\includegraphics[width=\textwidth]{images/retargeting/DCIL_results.jpg}
%\caption{\textbf{Image retargeting (DCIL).} The input images are shown in the red colored frame.}\label{fig: retargeting}
%\end{figure*}
%Super-Resolution (SR) aims to enhance the image quality and generate a \textit{high-resolution} (HR) image $I^H \in \mathbb{R}^{mt\times nt \times 3}$ given a \textit{low-resolution} (LR) image  $\hat{I}\in \mathbb{R}^{m\times n \times 3}$ and a scaling factor $t$. 
%%%%%%%%%%%%%%%%%%%%%%%%%%%%%%%%%%%%%%%%%%%%%%%%%%%%%%%%%%%%%%%%%%%%%%%%
\subsection{Super-Resolution.} \label{SSEC: super} 
CNN based SR has been studied in two ways. First, we can use pixel-to-pixel loss, which leads to high PSNR at the price of low perceptual quality \cite{zhang2018unreasonable, blau2018perception}. Second, we can use feature space loss or an adversarial loss to achieve higher perceptual quality \cite{ledig2016photo, johnson2016perceptual}. CL combines the two training data based approaches above to generate natural-looking images, with good structural similarity \cite{mechrez2018learning}.

The generator and the discriminator for SR are similar to the ones used in DSR. The loss function for SR is similar to DSR given in Eq.~\ref{eq: multiL} but without TV norm as there is no noise in the input images. 

%$\mathcal{G}_2$ for SR is similar to $\mathcal{G}_1$ defined in Eq.~\ref{eq: superN}. The discriminator is similar to the one defined in Eq.~\ref{eq: discriminatorSum} and Eq.~\ref{eq: discriminator}. 
%The description of loss terms is similar to the one provided in Eq.~\ref{eq: multiL}. %but without TV norm as there is no noise in the input image. 
%The loss function for SR is given in Eq.~\ref{eq: superL}.
%It helps the network to regenerate images at multiple resolutions by minimizing the difference between the generated outputs and the downsampled corrupted images (Fig.~\ref{fig: MEDS}). 
%\begin{equation}\label{eq: superL}
%\begin{split}
%\mathcal{L}_2 = \lambda_{\mathcal{C}} \; \mathcal{L}_{CL}({\mathcal{G}_1}(x), y) + & \lambda_{\mathcal{G}_1} \; \mathcal{L}_{GAN}({\mathcal{G}_1},{\mathcal{D}},\hat{I},x) \\ + & \lambda_{\mathcal{R}} \; \| {\mathcal{G}_1}(x) - U_t(\hat{I})) \|
%\end{split}
%\end{equation}

We perform 4$\times$SR on BSD100 data set. Fig.~\ref{fig: super} shows the perceptual quality comparison for 4$\times$SR. The average SSIM scores on BSD100 dataset are as follows. Mechrez et al. \cite{mechrez2018learning}: 0.64, ZSSR \cite{shocher2018zero}: 0.72, DIP \cite{Ulyanov2018CVPR}: 0.79, and our DCIL: 0.76. We found that MSE based methods capture strong prior for SR compared to the contextual loss and adversarial loss-based frameworks, which is counter-intuitive \cite{Ulyanov2018CVPR}. We confirm this by the following results (our run) \footnote{For BSD100, we have used SSIM values provided in \cite{mechrez2018learning}. For Set-5 and Set-14 dataset, we have used unsupervised implementation of  \cite{mechrez2018learning} for a fair comparison. SSIM values for \cite{Ulyanov2018CVPR} is computed by our run.}. For Set-5, the score are, Mechrez et al. \cite{mechrez2018learning}: 0.86, ZSSR \cite{shocher2018zero}: 0.87, DIP \cite{Ulyanov2018CVPR}: 0.90, and our DCIL: 0.87. For Set-14, Mechrez et al. \cite{mechrez2018learning}: 0.78, ZSSR \cite{shocher2018zero}: 0.76, DIP \cite{Ulyanov2018CVPR}: 0.81, and our DCIL: 0.79. Our interpretation of this phenomenon is as follows. Pixel-to-pixel comparison is converging to better optima in SR. However, in the case of DSR, a pixel-to-pixel comparison could be over-learning noise with features (Table~\ref{tab: multi}).  We believe that the performance of DIP and DCIL could probably be further improvised using hyper-parameter search. 

\begin{figure}\begin{center}
\resizebox{1.02\linewidth}{!}{%
\begin{subfigure}[b]{0.24\linewidth}\captionsetup{justification=centering}\begin{center}
\begin{minipage}{0.5\linewidth}
\includegraphics[width=\linewidth]{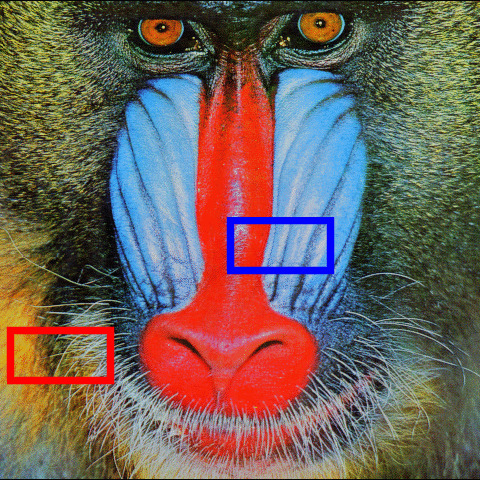}%
\end{minipage}%
\begin{minipage}{0.5\linewidth}
\includegraphics[width=\linewidth]{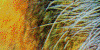} \\
\includegraphics[width=\linewidth]{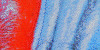}
\end{minipage}
\end{center}  \vspace*{-0.2cm} \caption{HR image} \end{subfigure}
\begin{subfigure}[b]{0.24\linewidth}\captionsetup{justification=centering}\begin{center}
\begin{minipage}{0.5\linewidth}
\includegraphics[width=\linewidth]{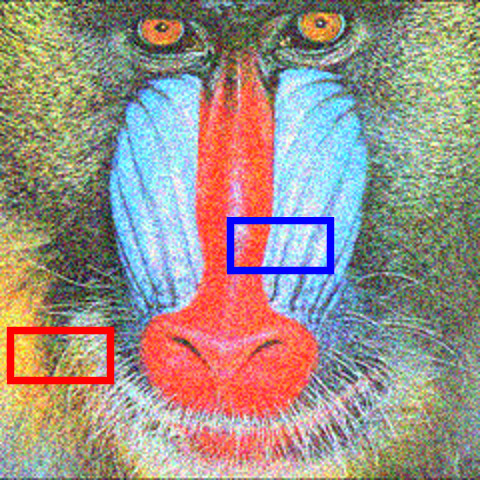} \end{minipage}%
\begin{minipage}{0.5\linewidth}
\includegraphics[width=\linewidth]{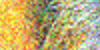} \\
\includegraphics[width=\linewidth]{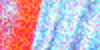}
\end{minipage}
\end{center}  \vspace*{-0.2cm} \caption{Corrupted} \end{subfigure}
\begin{subfigure}[b]{0.24\linewidth}\captionsetup{justification=centering}\begin{center}
\begin{minipage}{0.5\linewidth}
\includegraphics[width=\linewidth]{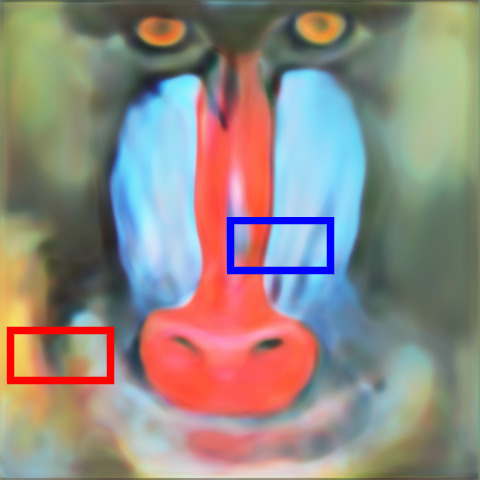} \end{minipage}%
\begin{minipage}{0.5\linewidth}
\includegraphics[width=\linewidth]{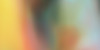}\\
\includegraphics[width=\linewidth]{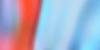}
\end{minipage}
\end{center}  \vspace*{-0.2cm} \caption{$\mathcal{G}_{noSkip}$} \end{subfigure}
\begin{subfigure}[b]{0.24\linewidth}\captionsetup{justification=centering}\begin{center}
\begin{minipage}{0.5\linewidth}
\includegraphics[width=\linewidth]{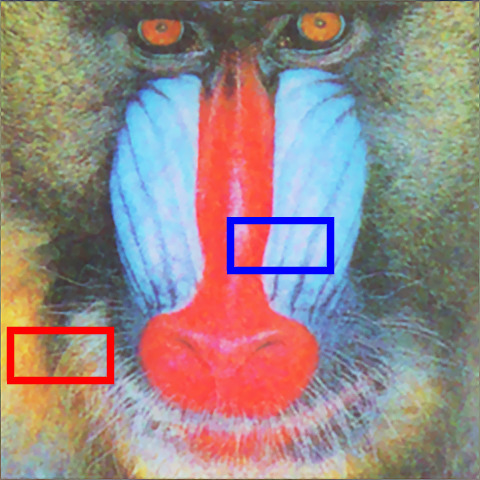} \end{minipage}%
\begin{minipage}{0.5\linewidth}
\includegraphics[width=\linewidth]{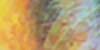}\\
\includegraphics[width=\linewidth]{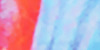}
\end{minipage}
\end{center}  \vspace*{-0.2cm} \caption{$\mathcal{G}_{skip}$} \end{subfigure}} \end{center} \vspace*{-0.5cm}
\caption{\textbf{Ablation Study.} 2$\times$Denoising-Super-Resolution with noise value $\sigma=100$. The network with skip connections $\mathcal{G}_{skip}$ performed better than the network without skip connections $\mathcal{G}_{noSkip}$. The above experiment study skip connections for multiple corruptions, unlike \cite{Ulyanov2018CVPR, mastan2019multi} (the images are best viewed after zooming). } \label{fig: ablationNetwork}
\end{figure}
%Here, $\mathcal{G}_{skip} = ( \{\phi_l \}_{l=1}^{10}, \{(i, 10-i)\}_{i=2}^{4}, \{ \}, \{ \})$ and $\mathcal{G}_{noSkip} = ( \{\phi_l \}_{l=1}^{10},  \{ \}, \{ \}, \{ \} )$. 
\begin{figure}
\begin{center}
\resizebox{\linewidth}{!}{%
\begin{subfigure}[b]{0.2\linewidth}
\includegraphics[width=\linewidth]{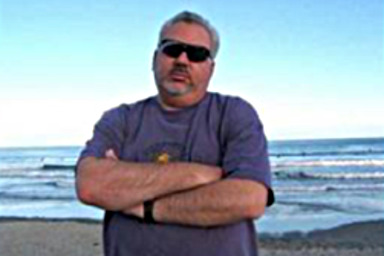}\caption{Input}
\end{subfigure}
\begin{subfigure}[b]{0.25\linewidth}
\includegraphics[width=\linewidth]{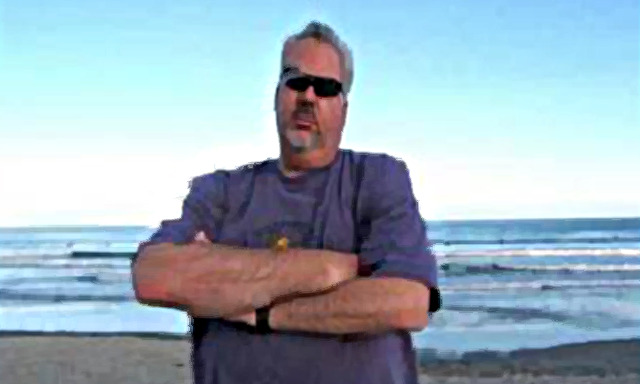}\caption{SC  \cite{avidan2007seam}} 
\end{subfigure} 
\begin{subfigure}[b]{0.25\linewidth}
\includegraphics[width=\linewidth]{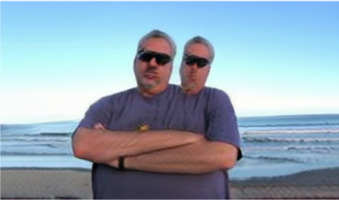}\caption{InGAN \cite{shocher2018internal}}
\end{subfigure}
\begin{subfigure}[b]{0.33\linewidth}
\includegraphics[width=\linewidth]{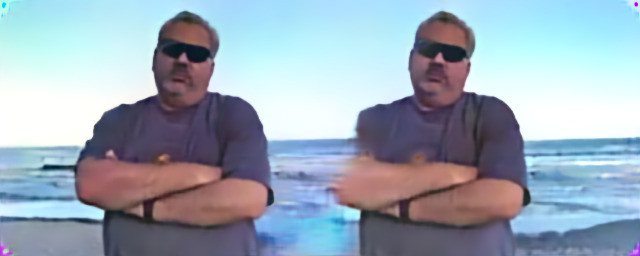}\caption{DCIL}
\end{subfigure}} \end{center} 
\vspace*{-0.5cm} \caption{\textbf{Failure Example.} The aim is to preserve the object context when performing image retargeting. SC \cite{avidan2007seam} deforms the object (\textit{i.e.}, man). InGAN does not partition the object well \cite{shocher2018internal}. DCIL partition the object, but the image feature of the elbow is not well-formed (the images are best viewed after zooming).}\label{fig: failureEx}
\end{figure}

%%%%%%%%%%%%%%%%%%%%%%%%%%%%%%%%%%%%%%%%%%%%%%%%%%%%%%%%%%%%%%%%%%%%%%%
%\begin{figure}[h] \centering
%\includegraphics[width=0.7\linewidth]{images/failureEx/ObjectAlignment.jpg}
%\caption{\textbf{Ablation study II.} Preservation of object features (alignment) for image retargeting.}\label{fig: RetargetingAblation}
%\end{figure}
%%%%%%%%%%%%%%%%%%%%%%%%%%%%%%%%%%%%%%%%%%%%%%%%%%%%%%%%%%%%%%%%%%%%%%%
\subsection{Image Retargeting.} \label{SSEC: retargeting} 

It is a content-aware image resizing operation which aims to output image with a different size, smaller or larger, and with a different aspect ratio. Image retargeting is performed in various ways. There are methods which are aimed to preserve only the salient objects and discarding/extending the object background  (e.g., \cite{cho2017weakly, wolf2007non}). Other methods (e.g., \cite{shocher2018internal, simakov2008summarizing} ) including our DCIL preserve the local sizes/aspect-ratios of the local objects while resizing the image.  The replication/reduction of the objects is desired to fill the scene with similar image features. 

Suppose input image $x$ is of size $(h,w)$. The scaling factors $s_h$ and $s_w$ are used as the input for retargeting. The retargeting objective is to output image $y$ with the size $(s_h h, s_w w)$. The description of the generator network for image retargeting is given in Eq.~\ref{eq: retargetingN}.
\begin{equation}\label{eq: retargetingN}
{\mathcal{G}_3} = \Big( \{G_l \}_{l=1}^{10}, \{(i, 10-i)\}_{i=2}^{4}, \{ \}, \{6\} \Big)
\end{equation}
Here, $\{G_l \}_{l=1}^{10}$ are the layers of the depth-5 encoder-decoder network. The network is equipped with skip connections from the layers of the encoder to the layers of the decoder. There are six residual blocks. The discriminator network is similar to the one defined in the Eq.~\ref{eq: discriminator}. 

The loss function for image retargeting is given in Eq.~\ref{eq: retargetingL}.
%The generator $G$ performs the image retargeting and outputs image $y$ with the size $(s_h h, s_w w)$. Next, the output of the generator $y$ is retargeted to the size of $x$ to compute the reconstruction loss (i.e., MSE). 
\begin{equation}\label{eq: retargetingL} 
\begin{split}
\mathcal{L}_3 = \lambda_{\mathcal{C}} \; \mathcal{L}_{CL}({\mathcal{G}_3}(x), y) + &\lambda_{GAN} \; \mathcal{L}_{GAN}({\mathcal{G}_3},{\mathcal{D}},x,y) \\ & + \lambda_{\mathcal{R}} \; \| {\mathcal{G}_3}({\mathcal{G}_3}(x)) - y \|
\end{split}
\end{equation}
 Here, $ \lambda_{\mathcal{C}}$, $\lambda_{GAN}$, and $\lambda_{\mathcal{R}}$ are the scaling factors. The adversarial loss $\mathcal{L}_{GAN}$ and the contextual loss $\mathcal{L}_{CL}$ both matches the distribution of image patch of the source image and the target images. Distribution matching is the essential requirement for image retargeting \cite{shocher2018internal}. Also, they both work for the non-aligned image data of the source and the target images, unlike DIP \cite{Ulyanov2018CVPR}.
 
We compute an automorphism as the cycle consistency check to preserve all the object details in the synthesized output \cite{shocher2018internal}. The automorphism retargets the generator output back to its source domain. Then we could perform the pixel comparison using reconstruction loss. It preserves the global image features in the retargeted image.

Our DCIL uses contextual learning to preserve the object features and object alignment in the image retargeting output better than Seam-Carving (SC) \cite{avidan2007seam} as shown in Fig.~\ref{fig: objStructure} and Fig.~\ref{fig: objAlign}. DCIL maximizes the feature learning and performs comparably to InGAN (Fig.~\ref{fig: failureEx}). 

%Seam carving provides good results \cite{avidan2007seam}. 
%Bi-directional similarity ensures the preservation of the local patches, but it could miss the relative location of the image patches \cite{simakov2008summarizing}. InGAN improvises Bi-directional similarity to preserve the local and the global structure \cite{shocher2018internal}. 
%In Fig.~\ref{fig: failure}, we show the limitations of contextual learning. % One way is to use the segmentation as the side information to improvise the supervision of the image features. We propose it as future work. 
\begin{table}[h]\setlength\extrarowheight{2pt}
\centering \resizebox{\linewidth}{!}{%
\begin{tabular}{|c|c|c|c|c|c|} \hline
& ZSSR \cite{shocher2018zero} & CL \cite{mechrez2018learning} & DIP \cite{Ulyanov2018CVPR} & InGAN \cite{shocher2018internal} & \textbf{DCIL (ours)}  \\ \hline
DSR & \xmark & \cmark & \cmark & \xmark & {\color{blue} \cmark}  \\ \hline
SR & \cmark & \cmark & \cmark & \cmark & {\color{blue} \cmark}  \\ \hline
Retargeting  & \xmark  & \xmark  & \xmark  & \cmark  & {\color{blue} \cmark}  \\ \hline 
\end{tabular}
}  \vspace*{-0.1cm} \caption{The table shows the comparison between various frameworks. DCIL (ours) is a generalized framework which performs all the tasks and generates images comparable to the other methods. We provide the extended version of Table 2 and the implementation details of DCIL in the supplementary material.}\label{table: compare}
\end{table}
%  ZSSR  \cite{shocher2018zero}, DIP \cite{Ulyanov2018CVPR}, InGAN \cite{shocher2018internal}, and ours DCIL do not use training data to learn the image prior unlike \cite{mechrez2018learning}. 

\noindent \textbf{User study.} We conducted a user study to evaluate the image retargeting results. We collected feedback from 58 human experts with a total of 290 votes. Each subject is asked to vote the perceptually better images constrained to the preservation of object properties. The percentage of the votes for SC \cite{avidan2007seam} is 35\%. Our DCIL got 65\% votes. The user study shows that DCIL performs good image retargeting. 

%We display output images with the reference image side-by-side on a webpage in the random order. It could be an interesting direction to develop a quality metric for the problems shown in the manuscript.
\section{Ablation Study and Limitations} \label{sec: ablation}
The limitations of the DCIL framework are due to the lack of contextual understanding by feature learning from a single image. 
Fig.~\ref{fig: ablationNetwork} shows that the network with skip connection outperforms the network without skip connections. Therefore, one needs to carefully design an application-specific network to maximize feature learning \cite{mastan2019multi}.  
Fig.~\ref{fig: failureEx} shows that contextual feature learning of DCIL leverages adversarial learning of InGAN for object partitioning limitations in image retargeting. However, the perceptual quality in the presence of object replications could be further improvised. 

%\section{Implementation Details}
%We have created a Tensorflow implementation of the DCIL framework. We use ADAM optimizer, and the batch size is `1' as there is only a single image as input. For image retargeting, DCIL input is cropped to make both height and width to be multiples of 128, and it outputs multiples of 128 (InGAN retarget to any size \cite{shocher2018internal}). To minimize the number of iterations (24K iterations), we use deep photo style transfer \cite{luan2017deep} (1K iterations) to transfer image features to retargeting output. For image restoration, the input is a cropped to multiples of 32, and we perform 1K iterations. Further details are provided in the supplementary material. 
\begin{figure} \centering
\begin{subfigure}[b]{0.22\linewidth}
\includegraphics[width=\linewidth]{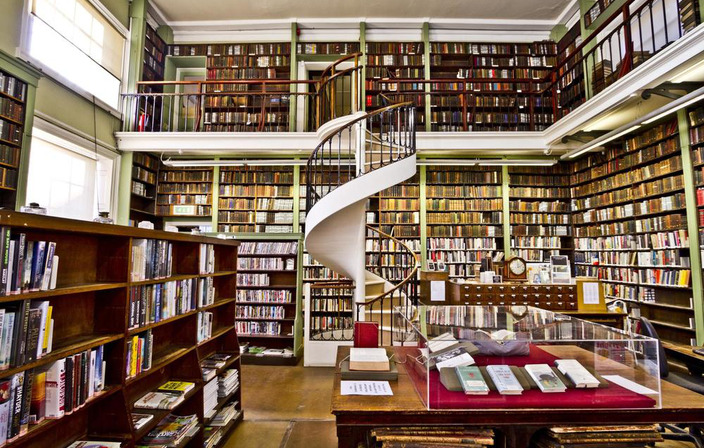}\caption{Image}
\end{subfigure}
\begin{subfigure}[b]{0.22\linewidth}
\includegraphics[width=\linewidth]{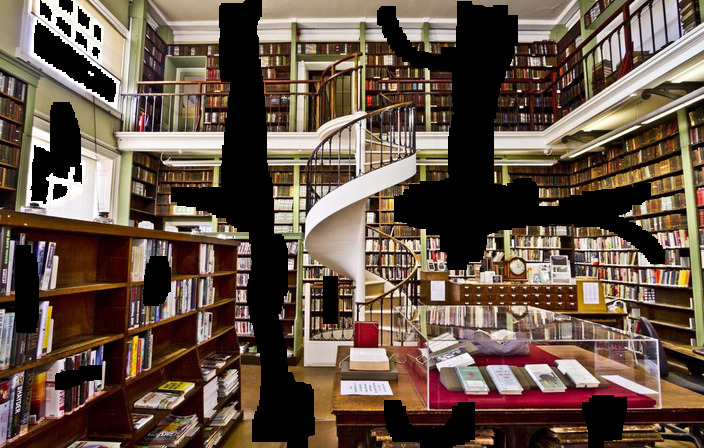}\caption{Mask}
\end{subfigure}
\begin{subfigure}[b]{0.22\linewidth}
\includegraphics[width=\linewidth]{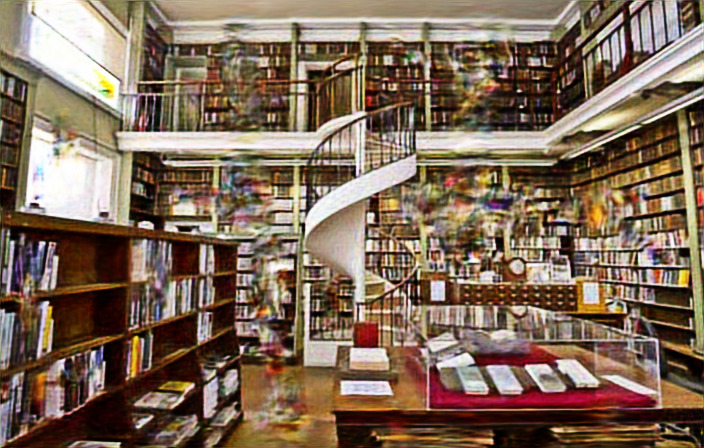}\caption{DIP \cite{Ulyanov2018CVPR}}
\end{subfigure}
\begin{subfigure}[b]{0.22\linewidth}
\includegraphics[width=\linewidth]{images/Inpainting/wacv_inpainting.jpg}\caption{DCIL}
\end{subfigure}  \vspace*{-0.2cm}
\caption{\textbf{Image Inpainting.} This shows the results for region inpainting (the images are best viewed after zooming).} \label{fig: inpainting}
\end{figure} 
\section{Discussion}\label{sec: discussion} 
%%%%%%%%%%%%%%%%%%%% WACV %%%%%%%%%%%%%%%%%%% 
%We perform image resize in three scenarios: Super-resolution (SR), Denoising-Super-Resolution (DSR), and content-aware Image Retargeting. SR and DSR are the image restoration problems which could be solved using the self-similarity prior based methods \cite{elad2006image, irani2009super, shocher2018zero}. Image restoration using DCIL is more related to DIP proposed by Ulyanov  \textit{et al.} \cite{Ulyanov2018CVPR}. We classify the image retargeting methods into the following types. Seam carving \cite{shamir2009seam, han2010optimal, frankovich2011enhanced}, warping based methods \cite{liu2005automatic, wang2008optimized, zhang2009shape, guo2009image, jin2010nonhomogeneous}, and CNN based methods \cite{cho2017weakly, shocher2018internal}. Image retargeting using DCIL is related to training data-independent InGAN \cite{shocher2018internal}.
%%%%%%%%%%%%%%%%%%%% WACV %%%%%%%%%%%%%%%%%%% 
DCIL is completely unsupervised and does not use training samples. It is different than the supervised methods RCAN \cite{zhang2018image} and DRLN \cite{anwar2019densely}, which use training data to perform image restoration. DCIL exploits the inherent self-similarity present in the source image. Ulyanov \textit{et al.} have shown that self-similarity prior emerged because of the convolutional operations tend to impose self-similarity in the generated images  \cite{Ulyanov2018CVPR}. DCIL incorporates image prior using the network structure implicitly. Similar to the DSR, SR, and image retargeting, it could also perform image inpainting (Fig.~\ref{fig: inpainting}). It is due to the self-similarity prior captured by DCIL helps to perform inpainting task. The quality of the deep prior for the various tasks depends upon the learning procedure. The network initially learns the image feature, but then it tends to over-learn the noise from the corrupted input \cite{mastan2019multi}. The learning procedure is generally more tricky when we perform distribution matching using GAN and CL. However, an exhaustive hyper-parameter search helped us in the above scenario. 
%It provides image features to the target image by optimizing feature learning from the single source image. DCIL performs image restoration without explicitly modeling the degradation process. It integrates contextual loss \cite{mechrez2018learning} with adversarial loss \cite{shocher2018internal} for image retargeting.  The essential property of the methods is the self-similarity present in the natural images. 

\section{Conclusion}
DCIL fits a randomly-initialized untrained generator. The structure of the network and the loss function are the main tools for unsupervised approaches described in the paper. We performed image resizing in many challenging scenarios. The performance depends upon the high correlation between the features of the source and the target images. For example, in the presence of high corruption due to noise in the source image, the performance of various methods degrade. We believe that it would be interesting to investigate the image statistics captured by DCIL for the other single image applications, \textit{e.g.}, image inpainting.

%\noindent \textit{Objects partitioning:} We do not know the way to perform the object partitioning. 

%%%%%%%%%%%%%%%%%%%%%%%%%%%%%%%%%%%%%%%%%%%%%%%%%%%%%%%%%%%%%%%%%%%%%%%
%\begin{figure}
%\includegraphics[width=0.10\textwidth]{images/jigsaw/blue.jpeg}
%\caption{Jigsaw Puzzle}\label{fig: jigsaw}
%\end{figure}
%%%%%%%%%%%%%%%%%%%%%%%%%%%%%%%%%%%%%%%%%%%%%%%%%%%%%%%%%%%%%%%%%%%%%%%

{\small
\bibliographystyle{ieee}
\bibliography{egbib}
}

\end{document}